\documentclass{iaus}
\usepackage{graphicx}

\title[Milky-Way Satellites] %% give here short title %%
{Modeling the Milky-Way Satellite galaxies}

\author[X. Kang]   %% give here short author list %%
{X. Kang$^1$}

\affiliation{$^1$Max-Planck-Institute for Astronomy, K\"onigstuhl 17, 
D-69117, Heidelberg, Germany\\email: {\tt kang@mpia.de}}

\pubyear{2008}
\volume{254}  %% insert here IAU Symposium No.
\pagerange{xxx-xxx}
\jname{The Galaxy Disk in Cosmological Context}
\editors{J. Andersen, J. Bland-Hawthorn \& B. Nordstr\"om}
\begin{document}

\maketitle

\begin{abstract}
  We revisit  the Milky Way satellite problem  using a semi-analytical
  model  of  galaxy formation  and  compare  the predicted  luminosity
  function   to   recent   result   from  the   SDSS.    With   cosmic
  photoionization, the  luminosity function can be  brought into broad
  agreement with  the data between $-15< M_{V}  <-2$. This improvement
  over  previous semi-analytical  model results  (e.g., Benson  et al.
  2002)  is from  our adoption  of improved  models for  galaxy merger
  history and  galaxy merging time-scales.  The  very faint satellites
  ($M_{v}  >  -5$)  formed  in  halos  with  virial  temperature  over
  $10^{4}K$ (  mass around $10^{9} M_{\odot}$ before accretion),
  but    their   baryon   content    are   strongly    suppressed   by
  photoionization. We  model the mass  evolution of the  subhalos, and
  compare the  predicted mass-to-light ratio  with the data.   We find
  that the measured total mass inside the luminous radii of satellites
  are about 5\% of their present total dark matter mass.


\end{abstract}

\firstsection % if your document starts with a section,
\section{Introduction}

One of  the most mysterious of  our Milky-Way Galaxy  is the ``missing
satellite problem'' that cold dark matter scenario predicts its number
of  satellite galaxies  to  be  around hundred,  but  only dozens  are
observed  (e.g.  Moore  et  al.   1999).  A  few  solutions have  been
proposed for  this contradictory.  One  instant question is if  we are
comparing the  right things, i.e,  how to relate the  observed stellar
velocity  dispersion to  the  measured circular  velocity from  N-body
simulation (Hayashi  et al.  1999). Another possible  solution is that
photoionization suppress  the gas accretion  in small halos,  and only
halos  formed  before  reionization  can form  stars.   (e.g.   Gnedin
2000). Also there is worry  about the incompleteness of observation as
very faint satellites have too  low surface brightness to be observed.
In recent years,  more faint satellites are observed  along with their
kinematics  and  mass   measurements,  and  the  satellite  luminosity
function is also well determined from the SDSS (Koposov et al.  2008).
One  the  other  hand,  theoretical  modelling  of  dark  matter  halo
formation history  (Cole et al.  2008) and galaxy  merging time-scales
(Jiang et  al. 2008,  Boylan-Kolchin et al.   2008) are  also improved.
Given  these  progress,  it  is  deserved  to  revisit  the  ``missing
satellite problem'' and  there have been a few  papers to address this
(e.g. Simon \& Geha 2007).

Here we  use the  semi-analytical model of  galaxy formation  (Kang et
al. 2005) to predict the luminosity function, mass-to-light ratios for 
satellite galaxies and compare them with the data.

\section{Model}

One of the  main ingredients to model the  satellite galaxy population
is to produce  their formation and assembly history.   Here we use the
Monte-Carlo merger tree  code from Parkinson et al.   (2008) to obtain
the  formation history  of a  Milk-Way  type galaxy  with mass  around
$10^{12}M_{\odot}$.  This new Monte-Carlo  algorithm is still based on
the  EPS formula  , but  revised  to match  the Millennium  Simulation
(Springel et al.  2005) results at different mass scales.  Cole et al.
(2008) have  shown that for halo with  mass around $10^{12}M_{\odot}$,
the merger tree  from the previous Monte-Carlo algorithm  (Cole et al.
2000)  produces too  strong evolution  and too  many major  mergers at
lower  redshift.  We  produce 1000  realizations of  the  merger trees
using  the new  Monte-Carlo  algorithm,  and in  Fig.1  we show  their
formation history  with comparisons  to the N-body  simulation results
(Stewart et  al.  2008, Giocoli  et al.  2008).   It can be  seen that
both the  evolution of the main  progenitors and the  mass function of
accreted subhalos agree well with the simulation results.
\begin{figure}
% \vspace*{-2.0 cm}
\begin{center}
\includegraphics[width=1.0\textwidth]{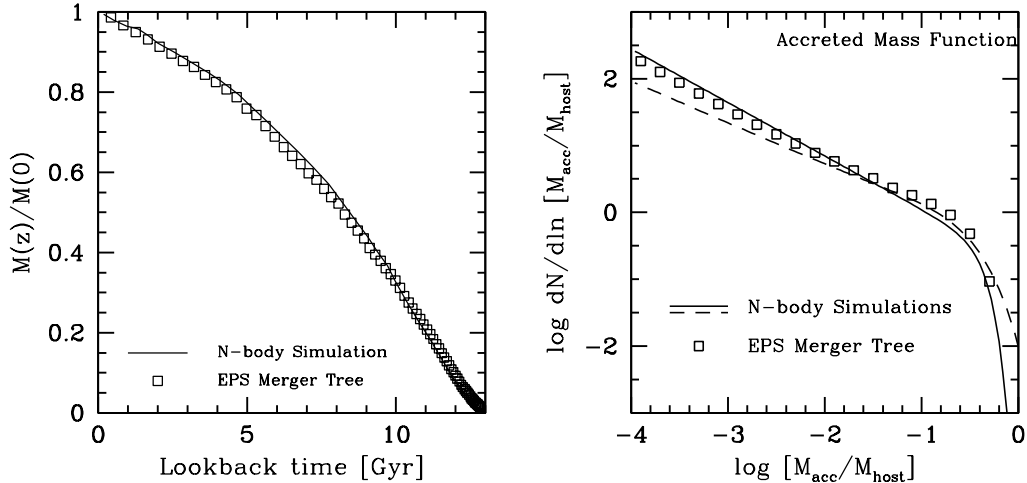}
% \vspace*{-1.0 cm}
\caption{The    formation   history    of   halos    with    mass   of
  $10^{12}M_{\odot}$ at  z=0.  Left panel:  the mass evolution  of the
  main  progenitors.  Right  panel:  the  mass  function  of  accreted
  subhalos by the main progenitors.   Good match are found with N-body
  simulations (Stewart et al. 2008, Giocoli et al. 2008).}
   \label{fig1}
\end{center}
\end{figure}

We then use  the semi-analytical model of Kang et  al. (2005) to model
the  formation of  satellite  galaxies along  the  merger trees.   The
semi-analytical  model  includes a  few  important physical  processes
governing galaxy  formation: hot gas cooling in  halos, star formation
from cold gas, supernova  feedback to reheat the inter-stellar medium,
stellar  evolution, galaxy  merger. We  update  the model  of Kang  et
al. (2005) by using an improved fitting formula for the galaxy merging
time-scales from Jiang et al.  (2008), who have shown that for massive
mergers, the survival time of  satellite galaxies in SPH simulation is
longer than the prediction from Lacey \& Cole (1993).

Here we also include a  simple model for photoionization from Kravtsov
et  al.  (2004).  In  case of  heating from  ionized photons,  the gas
content in halos  formed after reionization are suppressed  and can be
simply described by a filter  mass which increases with redshift.  The
filter  mass  increase  from  $10^{8}M_{\odot}$  at  z=8  to  $4\times
10^{10}M_{\odot}$ at  z=0 (Okamoto et  al. (2008) recently  argue that
the filter mass should be smaller).  The gas fraction in a filter mass
halo is  about half of  the universal fraction.   With photoionization
the  amount   of  gas  available  for  star   formation  is  decreased
significantly in less massive halos formed after reionization. In this
paper, we take the reionization redshift as z=7.

\section{Result}

\subsection{Luminosity function of satellites}

In Fig.2 we show the model luminosity function (LF) of satellites with
comparison to the  recent results of Koposov et  al.  (2008) from SDSS
DR5.  Koposov et  al measured the LF up to  $M_{V}=-2$, and found that
LF can  be described by a  power law with  slope of 0.1.  At  the very
faint  end  ($M_{V}  > -5$)  the  solid  circle  points in  Fig.2  are
extrapolated assuming  the satellite galaxies following  a NFW density
distribution,  and  empty  circles  are  assumed  with  an  isothermal
distribution (See Koposov  et al. 2008). It can be  seen that if there
is only  supernova feedback (dashed line), the  predicted total number
of  satellites are  more than  observed by  a factor  of 3.   With the
suppression of  gas accretion by photoionization, the  LF (solid line)
can be brought into abroad  agreement with the data.  This is expected
that the decrease of gas content produce less stellar mass.

Compared to the  model prediction of Benson et  al.  (2002), our model
produces more  luminous satellites  with $M_{V}<-15$. This  success is
credited to  the combination of  improved models for halo  merger tree
and  galaxy merging  time-scales. The  merger tree  used by  Benson et
al. (2002)  is based on Cole  et al.  (2000), which  produces too many
recent major mergers. As the  galaxy merging time is shorter for major
merger, so less is the number of survived massive satellites.  Also we
use  the new  fitting  formula from  Jiang  et al.  (2008) for  galaxy
merging  time-scales, which is  longer than  the often  used dynamical
friction time scale from Lacey \& Cole (1993).

\begin{figure}
%\vspace*{-0.4 cm}
\begin{minipage}{0.45\linewidth}
\centering
\includegraphics[width=6.0cm]{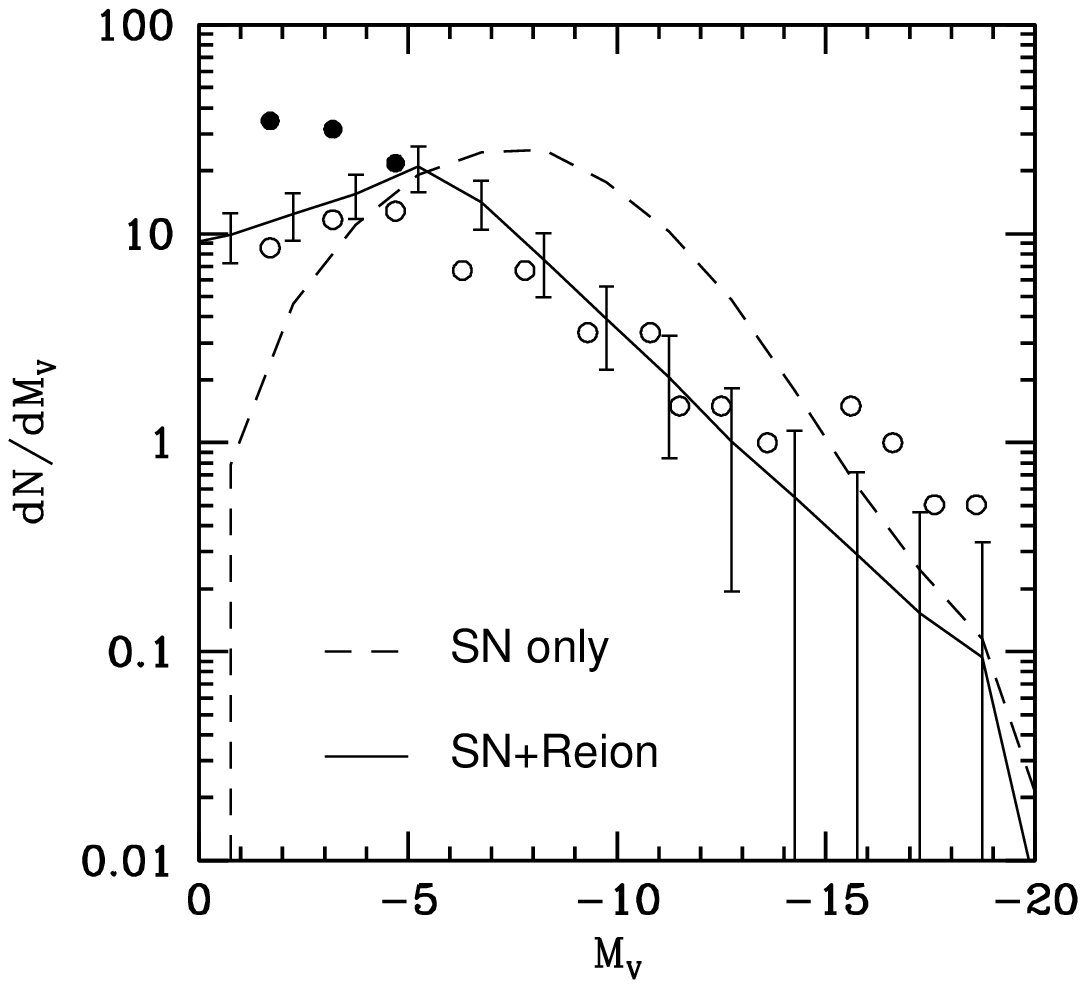}
% \vspace*{-1.0 cm}
\caption{The luminosity function of  satellite galaxies in a Milky-Way
  type   halo.   Data   points   (circles)   are   from   Koposov   et
  al.  (2008). Dashed  line is  model result  with  supernova feedback
  only,  and solid  line  (with  Poisson error)  is  model with  photo
  ionization included.}
\end{minipage}
\hspace{0.7cm}
%\vspace*{+0.7cm}
\begin{minipage}{0.45\linewidth}
\centering
\includegraphics[width=5.5cm]{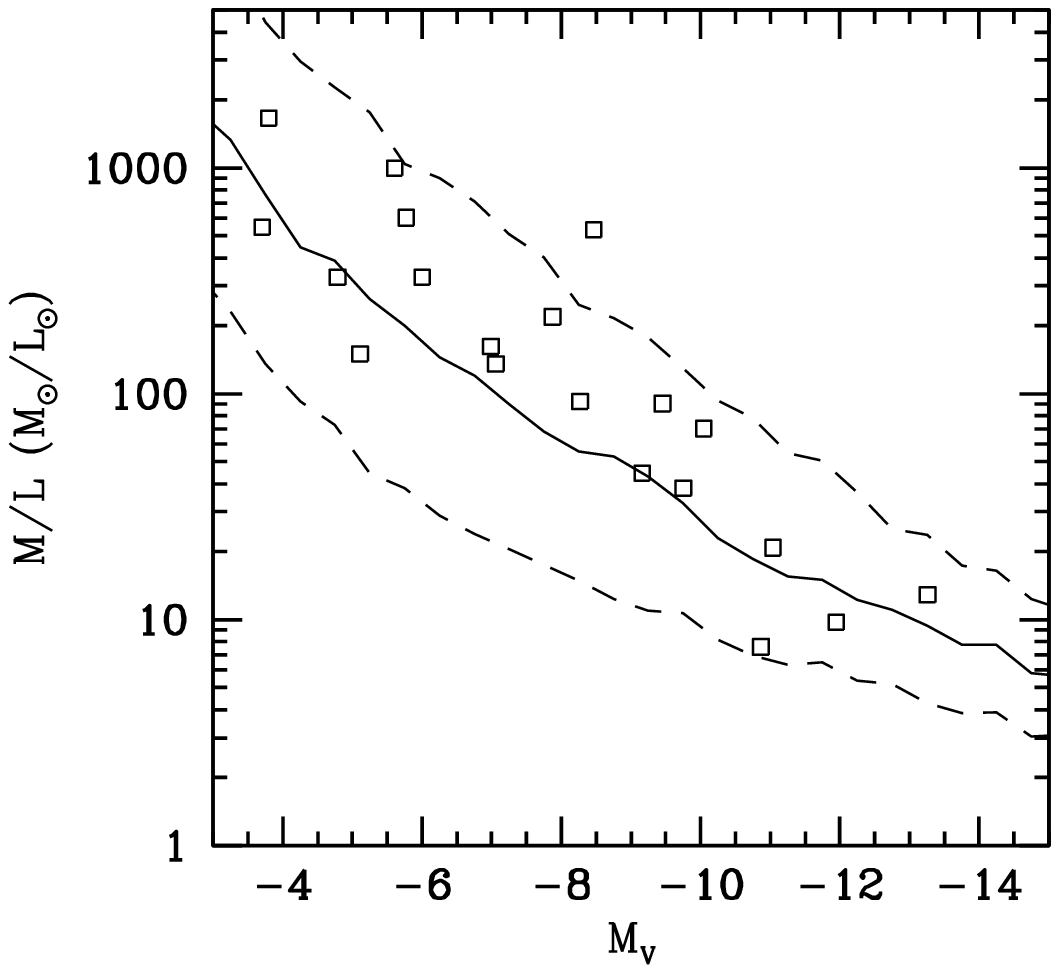}
\caption{The mass-to-light  ratio of satellite  galaxies.  Data points
  are from  compilation of  Simon \& Geha  (2007).  Solid line  is the
  model prediction with 20th  and 80th percentiles of the distribution
  (dashed  lines).  Here we  assume that  measured total  mass (inside
  luminous radii) of  satellites are 5\% of their  present dark matter
  mass.}
\end{minipage}
\end{figure}

As  we can  see that  without photoionization,  there are  only  a few
satellites fainter than  $M_{V}=-5$.  This is because hot  gas can not
cool via hydrogen line emission in halos with virial temperature below
$10^{4}K$  ($\sim  10^{9}M_{\odot}$)   and  $H_{2}$  cooling  is  very
inefficient. The  solid line shows that those  faint satellites formed
in halos with  virial temperature just over $10^{4}K$,  but have their
gas content strongly suppressed by photoionization (Similar conclusion
was also  obtained by  Kravtsov et  al.  2004).  In  our model,  it is
difficult to produce satellite  ($M_{V} \simeq -3$) with number around
30,  and this  favors the  satellites  to have  an isothermal  density
distribution.

\subsection{Mass-to-light ratio of satellites}

With the advent of  accurate measurements of satellites kinematics, it
is  possible to  obtain the  total mass  of satellite  galaxies inside
their luminous radii.  It is found that most  satellites are dominated
by  dark  matter  inside  their  luminous radii.   Here  we  show  the
mass-to-light ratio of satellites in  Fig.3, where the data points are
from compilation  by Simon \& Geha  (2007).  We model  the dark matter
mass evolution of satellite galaxies using the model of Giocoli et al.
(2008), and we further assume that  about 5\% of the total dark matter
mass  are inside  their luminous  radii. The  model  prediction (solid
line) along with 20th and 80th percentiles of the distribution (dashed
lines) are  shown in Fig.3. We  can find good agreement  with the data
from $M_{V} = -3$ up to $M_{V}=-15$.

\begin{figure}
% \vspace*{-2.0 cm}
\begin{center}
\includegraphics[width=0.5\textwidth]{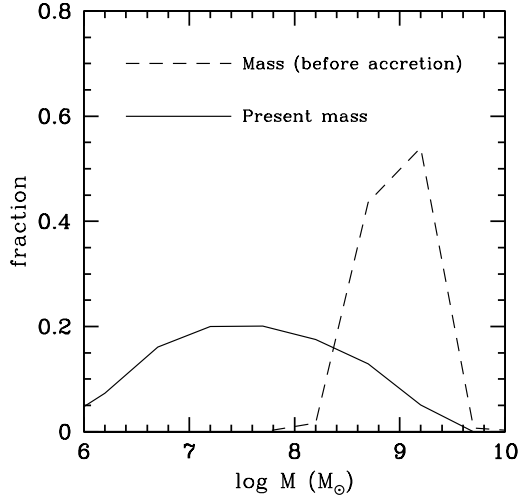}
% \vspace*{-1.0 cm}
\caption{The mass of faint satellites  ($M_{V} > -5$).  Dashed line is
  their  mass distribution  before accretion,  and solid  line  is the
  distribution   of   their  present   mass   after  evolution.    The
  distribution has a sharp peak at $10^{9}M_{\odot}$ before accretion,
  where hydrogen line emission cooling is efficient.  The present mass
  has a  broad distribution around $3\times  10^{7}M_{\odot}$, and the
  wide spread is from the dispersion of accretion times. }
   \label{fig4}
\end{center}
\end{figure}

The above  results of  LF and mass-to-light  ratio are  encouraging as
they indicates that we can model the luminosity and mass of satellites
simultaneously.  Now  we make prediction  for the dark matter  mass of
satellites faint than $M_{V} =-5$.   In Fig.4, we show their host halo
mass before accretion  (dashed line) and the present  dark matter mass
after evolution (solid line). As  we can see that the faint satellites
($M_{V} >  -5$) formed in  halos with mass peak  at $10^{9}M_{\odot}$,
and they have a broad distribution for their present mass with peak at
about $3\times  10^{7}M_{\odot}$.  The broad distribution  is from the
spread of their accretion times.

\section{Conclusion}

we revisit  the ``missing satellite problem''  using a semi-analytical
model of galaxy formation  combined with a high-resolution merger tree
from  Monte-Carlo algorithm (Parkinson  et al.   2008).  we  model the
luminosity function  and mass-to-light ratio for  the satellites.  The
model  luminosity function  agrees  well with  the  recent results  of
Koposov et al.   (2008) from the SDSS DR5  only if the photoionization
effect is included to suppress the gas fraction in less massive halos.
Our ability to produce  more luminous satellite galaxies than previous
semi-analytical  models  (e.g.,  Benson  et  al.  2002)  is  from  the
improvement  on  modelling of  halo  merger  tree  and galaxy  merging
time-scales.  Very  faint satellites  ($M_{V}>-5$) form in  halos with
virial temperature above $10^{4}K$, but their gas content are strongly
suppressed by photoionization. In addition their number density favors
an  isothermal density distribution  in the  Milky-Way.  We  model the
mass evolution of subhalos using  the model of Giocoli et al.  (2008),
and find  that the measured  total kinematic mass inside  the luminous
radii of satellite galaxies are about 5\% of their present dark matter
mass.

\section{Acknowledgements}

I thank Nicolas  Martin, Jeta T. de Jong and  Simon White for helpful
discussions.

\end{document}